\documentclass[aps,floats]{revtex4}
\usepackage{amsmath,amssymb}
\usepackage{graphicx,epsfig}
\usepackage[greek,english]{babel}
\usepackage{bbold}

\begin{document}
\bibliographystyle {plain}

\pdfoutput=1
\def\oppropto{\mathop{\propto}} 
\def\opsimeq{\mathop{\simeq}}
\def\opoverderline{\mathop{\overline}}
\def\operarrow{\mathop{\longrightarrow}}
\def\opsim{\mathop{\sim}}

\def\fig#1#2{\includegraphics[height=#1]{#2}}
\def\figx#1#2{\includegraphics[width=#1]{#2}}


\title{  Statistical physics of long dynamical trajectories \\
for a system in contact with several thermal reservoirs } 


\author{ C\'ecile Monthus }
 \affiliation{Institut de Physique Th\'{e}orique, 
Universit\'e Paris Saclay, CNRS, CEA,
91191 Gif-sur-Yvette, France}

\begin{abstract}
For a system in contact with several reservoirs $r$ at different inverse-temperatures $\beta_r$, we describe how the Markov jump dynamics with the generalized detailed balance condition can be analyzed via a statistical physics approach of dynamical trajectories $[{\cal C}(t)]_{0 \leq t \leq T}   $ over a long time interval $T \to + \infty$. The relevant intensive variables are the time-empirical density $\rho(\cal C)$, that measures the fractions of time spent in the various configurations ${\cal C}$, and   the time-empirical jump densities $k_r ({\cal C', \cal C})  $, that measure the frequencies of jumps from configuration ${\cal C}  $ to configuration ${\cal C '}  $ when it is the reservoir $r$ that furnishes or absorbs the corresponding energy difference ($E({\cal C '})- E({\cal C })$).

\end{abstract}

\maketitle

\section{ Introduction }

\label{sec_intro}

The statistical physics of equilibrium is based 
on the ergodic principle, that states the equivalence
between the 'time average' of any observable $O$
 over a sufficiently long time dynamical trajectory $[{\cal C}(t)]_{0 \leq t \leq T}  $
and an appropriate 'ensemble average' over some probability $P^{ensemble}_{eq}({\cal C})$
over the configurations ${\cal C}$ 
\begin{eqnarray}
\frac{1}{T} \int_0^{T} dt 
O \left(  {\cal C}(t)  \right)  
\opsimeq_{T \to \infty}
 \sum_{ {\cal C}} O \left(  {\cal C}  \right) 
P^{ensemble}_{eq}\left({\cal C}   \right)
\label{ergodic}
\end{eqnarray} 
The various equilibrium ensembles are adapted to the physical conditions one is interested in :
for instance, the microcanonical ensemble is appropriate for isolated systems conserving the total energy, 
while the canonical ensemble is tailored for systems in contact with a single thermal reservoir,
with relations between the two in the thermodynamic limit 
(see the short reminder in Appendix \ref{app_eq}).

For non-equilibrium systems, it is thus natural to also base the analysis on time-averages over long 
dynamical trajectories. However, it turns out that the time-empirical density
measures the fractions of time spent in the various configurations ${\cal C}$
 (direct analog of Eq. \ref{ergodic})
\begin{eqnarray}
 \rho( {\cal C} ) \equiv \frac{1}{T} \int_0^T dt \ \delta_{{\cal C}(t) , {\cal C} }
\label{rhoc}
\end{eqnarray}
is not sufficient, and has to supplemented by the empirical flows between configurations,
i.e. for instance the jump densities from configurations
${\cal C}  $ to configuration ${\cal C '}  $ for the case of jump processes
\begin{eqnarray}
 k({\cal C '} , {\cal C} ) \equiv \frac{1}{T}  
\sum_{ t \in [0,T] : {\cal C}(t^+) \ne {\cal C}(t^-)  } 
\delta_{{\cal C}(t^+) , {\cal C'} } \ \delta_{{\cal C}(t^-) , {\cal C} }
\label{kcc}
\end{eqnarray}
For Markovian dynamics, 
the joint probability of the empirical density and of the empirical flows
is known to follow a very general large deviation form with respect to the large time $T$
(the general theory of large deviations is summarized in the reviews \cite{oono,ellis,review_touchette} 
with applications to various fields).
In addition, the corresponding rate function (known as 'Level 2.5' in the classification of Large Deviations)
is explicit within various frameworks, namely 
 for discrete-time/discrete-space Markov chains \cite{fortelle_thesis,fortelle_chain,c_largedevdisorder,review_touchette},
for continuous-time/discrete-space Markov jump processes \cite{fortelle_thesis,fortelle_jump,maes_canonical,maes_onandbeyond,wynants_thesis,chetrite_formal,BFG1,BFG2}
as well as for continuous-time/continuous-space diffusion processes \cite{wynants_thesis,maes_diffusion,chetrite_formal,engel}.
Moreover, this 'Level 2.5' formulation allows to reconstruct any time-additive observable
of the dynamical trajectory via its decomposition in terms of the empirical densities and of the empirical flows,
and is thus closely related to the studies focusing on the generating functions of time-additive observables 
via deformed Markov operators, that have attracted a lot of interest recently
in various models
 \cite{derrida-lecture,lecomte_chaotic,lecomte_thermo,lecomte_formalism,
lecomte_glass,kristina1,kristina2,chetrite_canonical,chetrite_conditioned,lazarescu_companion,lazarescu_generic,touchette_langevin,derrida-conditioned,bertin-conditioned}.
More generally, the idea that the theory of large deviations is the appropriate language
to analyze non-equilibrium dynamics has been emphasized from various perspectives (see the reviews  \cite{derrida-lecture,harris_Schu,searles,harris,mft,lazarescu_companion,lazarescu_generic}
and the PhD Theses \cite{fortelle_thesis,vivien_thesis,chetrite_thesis,wynants_thesis} 
 or HDR Thesis \cite{chetrite_HDR}).

In the present paper, we focus on the Markov jump dynamics of a system in contact with $R>1$
thermal reservoirs $r=1,..,R$ as described by the generalized-detailed-balance condition.
We wish to keep track in the jump densities $k({\cal C '} , {\cal C} )$ from $\cal C $ to $\cal C' $ of Eq. \ref{kcc}
of the reservoir $r$ that is responsible for the jump,
i.e. of the reservoir $r$ that furnishes or absorbs the energy difference between the initial and final configuration of the system
\begin{eqnarray}
 k({\cal C', \cal C}) = \sum_{r=1}^R  k_r ({\cal C', \cal C}) 
\label{kr}
\end{eqnarray}
The goal is to extend the standard 'Level 2.5' large deviation framework mentioned above 
\cite{fortelle_thesis,fortelle_jump,maes_canonical,maes_onandbeyond,wynants_thesis,chetrite_formal,BFG1,BFG2}
to this finer analysis of the dynamical trajectories that contains all the information on the heats exchanges with the various reservoirs.

The paper is organized as follows.
In section \ref{sec_markov}, we recall the Markov jump dynamics framework
to describe a system in contact with several reservoirs, and 
we describe the relevant time-empirical observables of dynamical trajectories.
 In section \ref{sec_probatraj}, the probability of individual trajectories is analyzed
in order to stress the important notion of intensive action of a function of the relevant empirical observables.
Section \ref{sec_entropy} is devoted to the Boltzmann entropy
of the set of dynamical trajectories with given empirical observables.
In section \ref{sec_2.5}, the large deviations properties 
of the joint probability of the time-empirical density and jump densities
are discussed. Finally the link with the statistics of other
time-empirical additive observables is described in section \ref{sec_additive}.
Section \ref{sec_conclusion} summarizes our conclusions.
Appendix \ref{app_eq} contains a short summary of the statistical physics of equilibrium
in order to make more explicit the comparison with the equations of the text concerning non-equilibrium.

\section{ Markov jump dynamics for a system in contact with several reservoirs  }

\label{sec_markov}

\subsection{ Markov jump dynamics in configuration space }

Since our goal is to remain at the same level of generality as the statistical physics theory of Equilibrium
(see the short reminder in Appendix \ref{app_eq}),
we will consider that the system can be in a certain number of configurations $\cal C  $ of energies $E(\cal C)$.
The Markov jump dynamics between the configurations of the system
is defined by the transition rates $w({\cal C', \cal C})$ from $\cal C $ to $\cal C' \ne \cal C$
for the dynamical elementary moves ${\cal C'} \leftarrow {\cal C} $ that are allowed in the model
(for instance single-spin-flip dynamics in spin models).
It is convenient to introduce the following notation for the total rate out of configuration $\cal C $ 
\begin{eqnarray}
w^{out}( {\cal C}) \equiv \sum_{{\cal C'} \ne  \cal C }w({\cal C', \cal C})
\label{wtout}
\end{eqnarray}
Then the probability $P_t(\cal C) $ to be in configuration $\cal C $ at time $t$
then evolves with the master equation 
\begin{eqnarray}
\partial_t P_t(\cal C) && = \sum_{\cal C' \ne \cal C}  w({\cal C, \cal C'}) P_t({\cal C'}) - w^{out}({\cal C}) P_t({\cal C})
\label{master}
\end{eqnarray}
As recalled in the Introduction, many studies have been devoted to the analysis 
of dynamical trajectories within this general framework.

\subsection{ Generalized detailed balance}

Here we wish to focus on the case where the system is in contact with $R$ reservoirs $r=1,2,..,R$
characterized by different inverse temperature $\beta_r$.
We will consider the generalized-detailed-balance condition in the following form (see the review \cite{broeck} and references therein) :

(i) the transition rate $w({\cal C', \cal C}) $ is given by the sum over the contributions $ w_r ({\cal C', \cal C})  $
of the reservoirs that allow this transition 
 (i.e. when the reservoir $r$ is able to furnishes or absorb the energy difference between the initial and final configuration of the system, otherwise the rate is zero)
\begin{eqnarray}
 w({\cal C', \cal C}) = \sum_{r=1}^R  w_r ({\cal C', \cal C}) 
\label{genedetailed}
\end{eqnarray}
so that Eq. \ref{wtout} becomes
\begin{eqnarray}
w^{out}( {\cal C}) \equiv \sum_{{\cal C'} \ne  \cal C }  \sum_{r=1}^R  w_r ({\cal C', \cal C}) 
\label{wtoutr}
\end{eqnarray}

(ii) the transitions rates $w_r ({\cal C', \cal C})  $
associated to the reservoir $r$ satisfy the usual detailed-balance condition
at the inverse temperature $\beta_r$ of the reservoir $r$
\begin{eqnarray}
 w_r({\cal C, \cal C'}) e^{- \beta_r E({\cal C'})} =  w_r({\cal C', \cal C}) e^{- \beta_r E({\cal C})}
\label{detailedr}
\end{eqnarray}
In particular, if the reservoir $r$ were alone, the stationary state of the system
would be the canonical distribution at inverse temperature $\beta_r$.

Since Eq \ref{detailedr} only fixes the ratios 
\begin{eqnarray}
\frac{w_r({\cal C', \cal C}) } { w_r({\cal C, \cal C'}) } 
= e^{- \beta_r \left[ E({\cal C'})  -E({\cal C}) \right] } 
\label{detailedratio}
\end{eqnarray}
but not the products $[w_r({\cal C', \cal C})  w_r({\cal C, \cal C'}) ] $,
 various choices for the transition rates exist in the literature.
The simplest choice from a technical point of view 
corresponds to the case $[w_r({\cal C', \cal C})  w_r({\cal C, \cal C'}) ]=1 $ \cite{rates_simple1977,rates_simple2013,rates_simple2016}
\begin{eqnarray}
w^{simple}_r({\cal C', \cal C})  
= e^{- \frac{ \beta_r }{2} \left[ E({\cal C'})  -E({\cal C}) \right] }
\label{simple}
\end{eqnarray}
while a more physical choice (in order to have transition rates that remain finite in the zero-temperature limit $\beta_r=\infty$)
is given by the Glauber transition rates \cite{glauber}
\begin{eqnarray}
w^{Glauber}_r({\cal C', \cal C})  
= \frac{ e^{- \frac{ \beta_r }{2} \left[ E({\cal C'})  -E({\cal C}) \right] } }
{e^{\frac{ \beta_r }{2}\left[ E({\cal C'})  -E({\cal C}) \right] } + e^{- \frac{ \beta_r }{2} \left[ E({\cal C'})  -E({\cal C}) \right] } } 
= \frac{ e^{- \frac{ \beta_r }{2} \left[ E({\cal C'})  -E({\cal C}) \right] } }
{ 2 \cosh \left( \frac{\beta_r }{2}\left[ E({\cal C'})  -E({\cal C}) \right] \right) } 
\label{glauber}
\end{eqnarray}

\subsection{ Relevant time-empirical observables for the dynamical trajectories and their constraints }

\label{sub_empi}

As recalled in the Introduction, the fundamental time-empirical observables for a
Markov jump dynamical trajectory $[{\cal C}(t)]_{0 \leq t \leq T} $ over the large-time interval $T$
are \cite{fortelle_thesis,fortelle_jump,maes_canonical,maes_onandbeyond,wynants_thesis,chetrite_formal,BFG1,BFG2}

a) the time-empirical density $\rho( {\cal C} )$ of Eq. \ref{rhoc}
with the normalization
\begin{eqnarray}
\sum_{\cal C}  \rho( {\cal C} ) =1
\label{rhotc}
\end{eqnarray}

b) the time-empirical jump densities $k({\cal C '} , {\cal C} ) $ of Eq. \ref{kcc},
with the following constraint for each ${\cal C} $
\begin{eqnarray}
 \sum_{ \cal C ' } k({\cal C '} , {\cal C} ) = \sum_{ \cal C ' } k({\cal C } , {\cal C'} )
\label{kcconstraint}
\end{eqnarray}
This quantity represents the number of sojourns in configuration ${\cal C} $ over the time $T$,
as computed from the number of departures from ${\cal C} $
or from the number of arrivals into ${\cal C} $ (for large $T$, one may neglect the
boundary terms where $ {\cal C}$ happens to be the initial configuration at $t=0$
or the final configuration at $t=T$).

In the present paper, we keep track of the reservoir $r$ in the jump densities (Eq. \ref{kr})
so that it is convenient to introduce the following notation
to summarize the constraints of Eq. \ref{rhotc} and \ref{kcconstraint} with the decomposition of Eq. \ref{kr},
\begin{eqnarray}
{\cal D} \left[ \rho(. ),  k_.(.,. ) \right] \equiv \left[ \delta(1-\sum_{\cal C}  \rho( {\cal C} ) ) \right]
 \prod_{\cal C} \delta \left( \sum_{\cal C' \ne \cal C} \sum_{r=1}^R k_r({\cal C'}, {\cal C}) - \sum_{\cal C' \ne \cal C} \sum_{r=1}^R k_r({\cal C},{ \cal C'})  \right) 
\label{dconstraint}
\end{eqnarray}
(the notation $\delta(Y)$ represents the discrete Kronecker symbol $\delta_{0,Y}$
but has been chosen here for better readability of the argument $Y$).

\subsection{ Replacing the empirical jump densities $k_r({\cal C', \cal C}) $ by the empirical activities $a_r({\cal C', \cal C}) $ and 
currents $j_r({\cal C', \cal C}) $ }

\label{sub_aj}

For each given pair of configurations $ \cal C$ and $\cal C' $, 
one can choose some order $ \cal C < \cal C'$, 
and decide to replace the two time-empirical 
jump densities $k_r({\cal C', \cal C})$ and $k_r({\cal C, \cal C'})$ 
by their antisymmetric and symmetric parts called respectively 
the time-empirical current $j_r({\cal C', \cal C})  $ and the time-empirical
activity $a_r({\cal C', \cal C}) $
\cite{maes_onandbeyond}
\begin{eqnarray}
j_r({\cal C', \cal C}) && =k_r({\cal C', \cal C}) -  k_r({\cal C, \cal C'})= - j_r({\cal C, \cal C'})
\nonumber \\
a_r({\cal C', \cal C})  && = k_r({\cal C', \cal C}) +  k_r({\cal C, \cal C'})  =  a_r({\cal C, \cal C'})
\label{jafromq}
\end{eqnarray}

This decomposition is more appropriate when one wishes to analyze 
the properties upon time reversal \cite{maes_onandbeyond}:
if the dynamical trajectory $[{\cal C}(t)]_{0 \leq t \leq T} $
is characterized by the time empirical observables $(\rho({\cal C}),a_r({\cal C', \cal C}),j_r({\cal C', \cal C}))$, then the time-reversed trajectory $[{\cal C}^{Backward}(t)]_{0 \leq t \leq T} $
defined by
\begin{eqnarray}
{\cal C}^{Backward}(t) \equiv {\cal C}(T-t)
\label{backward}
\end{eqnarray}
is characterized by the same empirical density and empirical activities,
but by opposite empirical currents
\begin{eqnarray}
\rho^{Backward}({\cal C}) && =\rho({\cal C})
\nonumber \\
a^{Backward}_r({\cal C', \cal C})  && =a_r({\cal C, \cal C'}) = a_r({\cal C', \cal C})
\nonumber \\
j_r^{Backward}({\cal C', \cal C}) && = j_r({\cal C, \cal C'}) = - j_r({\cal C', \cal C})
\label{jabackward}
\end{eqnarray}

Via this change of variables, 
the constraints of Eq. \ref{dconstraint} do not contain the activities $a_r({\cal C', \cal C}) $,
i.e. they involve only the empirical density $\rho( {\cal C} ) $ and the empirical currents $j_r({\cal C', \cal C}) $
\begin{eqnarray}
{\cal D} \left[ \rho(. ),  j_.(.,. ) \right]
\equiv \left[ \delta(1-\sum_{\cal C}  \rho( {\cal C} ) ) \right]
 \prod_{\cal C} \delta \left( \sum_{\cal C' \ne \cal C} \sum_{r=1}^R j_r({\cal C'}, {\cal C})   \right) 
\label{dconstraintj}
\end{eqnarray}

\subsection{Link with thermodynamics : intensive time-empirical heats $q_r$ furnished by the various reservoirs  }

During the time interval $[0,T]$,
each reservoir $r$ will furnish some extensive empirical heat $(T q_r)$,
 either positive or negative,
where the intensive time-empirical heat $q_r$ reads
in terms of the empirical jump densities $k_r ({\cal C', \cal C})  $,
or in terms of the empirical currents $j_r ({\cal C', \cal C})  $ introduced above
\begin{eqnarray}
 q_r  && =  \sum_{\cal C}  \sum_{\cal C' \ne \cal C}  
 \left[E({\cal C'})- E({\cal C}) \right] k_r ({\cal C', \cal C}) 
\nonumber \\
&& = \sum_{\cal C}  \sum_{\cal C' \ne \cal C}  
 \left[E({\cal C'})- E({\cal C}) \right] \frac{ j_r ({\cal C', \cal C}) }{2}
= \sum_{\cal C}  \sum_{\cal C' > \cal C}  
 \left[E({\cal C'})- E({\cal C}) \right]  j_r ({\cal C', \cal C}) 
\label{qr}
\end{eqnarray}
Besides this decomposition into pairs of configurations,
it is useful to write the following
alternative expression with respect to the energy of a single configuration
\begin{eqnarray}
 q_r  && = -  \sum_{\cal C}   E({\cal C}) \left(  \sum_{\cal C' \ne \cal C}    j_r ({\cal C', \cal C}) \right)
\label{qre}
\end{eqnarray}
in order to make obvious that the sum over the reservoirs vanishes
as a consequence as the constraint of Eq. \ref{dconstraintj}
\begin{eqnarray}
\sum_{r=1}^R q_r  && 
= -  \sum_{\cal C}   E({\cal C}) \left(  \sum_{\cal C' \ne \cal C} \sum_{r=1}^R   j_r ({\cal C', \cal C}) \right) =0
\label{qrsum0}
\end{eqnarray}
as it should since the energy cannot accumulate in the system.
This is the empirical version of the first principle of thermodynamics at the level of a very long dynamical trajectory
(see the review \cite{broeck} and references therein).


\section{ Probabilities of dynamical trajectories : notion of intensive action }

\label{sec_probatraj}

\subsection{ Extensive action ${\cal A} \left( [{\cal C}(t)]_{0 \leq t \leq T} \right) $ of a dynamical trajectory $[{\cal C}(t)]_{0 \leq t \leq T} $  }

A dynamical trajectory ${\cal C}(t)$ on the time interval $0 \leq t \leq T$
corresponds to a certain number $M \geq 0 $ of jumps $m=1,..,M$ occuring at times $0<t_1<...<t_M<T$
between the successive configurations $({\cal C}_0 \to {\cal C}_1 \to {\cal C}_2.. \to {\cal C}_M)$ that are visited between these jumps. Since we wish to keep track of the reservoirs labels $r_m$
that are responsible of the jumps at times $t_m$, it will be convenient to denote
this trajectory as
\begin{eqnarray}
[{\cal C}(t)]_{0 \leq t \leq T} = \left[ {\cal C}_0; (t_1,r_1) ; {\cal C}_1; (t_2,r_2) ; {\cal C}_2;
(t_3,r_3) ;... ;
 {\cal C}_{M-1} ; (t_{M},r_M) ; {\cal C}_M \right]
\label{traject}
\end{eqnarray}

The probability of this trajectory conditioned to start in configuration ${\cal C}_0 $
reads in terms of the transitions rates 
\begin{eqnarray}
&& {\cal P} \left( 
[{\cal C}(t)]_{0 \leq t \leq T} 
 =  \left[ {\cal C}_0; (t_1,r_1) ; {\cal C}_1; (t_2,r_2) ; {\cal C}_2;(t_3,r_3) ;... ;
 {\cal C}_{M-1} ; (t_{M},r_M) ; {\cal C}_M \right]
 \right)
\nonumber \\
&& = 
 e^{- (T-t_M) w^{out}({\cal C}_M) } 
 w_{r_M}({\cal C}_M , {\cal C}_{M-1} ) 
 e^{- (t_M-t_{M-1} ) w^{out}({\cal C}_{M-1}) } 
w_{r_{M-1}}({\cal C}_{M-1} , {\cal C}_{M-2} ) ...
\nonumber \\ 
&&
  .... w_{r_2}({\cal C}_{2} , {\cal C}_{1} ) 
 e^{- (t_2-t_1) w^{out}({\cal C}_1) } 
 w_{r_1}({\cal C}_1 ,{\cal C}_{0} )
 e^{- t_1 w^{out}({\cal C}_0) } 
\label{ptraject}
\end{eqnarray}

The normalization over all possibles trajectories on $[0,T]$
for a fixed initial configuration ${\cal C}_0$ 
involves the sum over the number $M$ of jumps, the sum 
over the $M$ configurations $({\cal C}_1,...,{\cal C}_M$)
where ${\cal C}_m$ has to be different from ${\cal C}_{m-1}$, 
the sum over the possible reservoirs $r_m$ for each jump,
the integration over the jump times with the measure $dt_1... dt_M$ and the constraint $0<t_1<...<t_M<T $
\begin{eqnarray}
&& 1  =\sum_{M=0}^{+\infty}  \int_0^T dt_M \int_0^{t_M} dt_{M-1} ... \int_0^{t_2} dt_{1} 
\sum_{r_M=1}^R ... \sum_{r_2=1}^R \sum_{r_1=1}^R
\sum_{{\cal C}_1 \ne {\cal C}_0} \sum_{{\cal C}_2 \ne {\cal C}_1} 
... 
\sum_{{\cal C}_{M-1} \ne {\cal C}_{M-2}}  
\sum_{{\cal C}_M \ne {\cal C}_{M-1}} 
\nonumber \\
&&  {\cal P} \left( [{\cal C}(t)]_{0 \leq t \leq T}  
=  \left[ {\cal C}_0; (t_1,r_1) ; {\cal C}_1; (t_2,r_2) ; {\cal C}_2;... ;
 {\cal C}_{M-1} ; (t_{M},r_M) ; {\cal C}_M \right]
 \right)
\label{norma}
\end{eqnarray}

The trajectory probability of Eq. \ref{ptraject}
 can be rewritten
\begin{eqnarray}
  {\cal P} \left( [{\cal C}(t)]_{0 \leq t \leq T} \right) 
=  e^{- {\cal A} \left( [{\cal C}(t)]_{0 \leq t \leq T} \right)  }
\label{lnptraject}
\end{eqnarray}
in terms of the action
\begin{eqnarray}
  {\cal A} \left( [{\cal C}(t)]_{0 \leq t \leq T} \right) 
=   \int_0^T dt w^{out}({\cal C}(t))     - \sum_{m=1}^M
\ln \left[  w_{r_m}({\cal C}(t_m^+) , {\cal C}(t_m^-) )   \right]
\label{actiontraject}
\end{eqnarray}
while the normalization of Eq. \ref{norma} can be formally rewritten as
\begin{eqnarray}
&& 1  = \sum_{ [{\cal C}(t)]_{0 \leq t \leq T} } {\cal P} \left( [{\cal C}(t)]_{0 \leq t \leq T} \right) 
 = \sum_{ [{\cal C}(t)]_{0 \leq t \leq T} }   e^{- {\cal A} \left( [{\cal C}(t)]_{0 \leq t \leq T} \right)  }
\label{normaformal}
\end{eqnarray}
where the sum over all the possible trajectories $[{\cal C}(t)]_{0 \leq t \leq T}  $
is meant to replace the more explicit first line of Eq. \ref{norma}.

\subsection{ Intensive action $ A[ \rho(.), k_.(.,.) ] $ of a dynamical trajectory in terms of its intensive empirical observables  }

The action of the dynamical trajectory of Eq. \ref{actiontraject} is extensive with respect to the large time-interval $T$
\begin{eqnarray}
  {\cal A} \left( [{\cal C}(t)]_{0 \leq t \leq T} \right) 
&& = T A[ \rho(.), k_.(.,.) ]
\label{actionempiricalt}
\end{eqnarray}
and the corresponding intensive action $ A[ \rho(.), k_.(.,.) ] $ reads 
in terms of the empirical density $\rho({\cal C})$ and of the empirical jump densities
$k_r({\cal C '} , {\cal C} )$ of the trajectory
\begin{eqnarray}
 A[ \rho(.), k_.(.,.) ]
&& =\sum_{{\cal C} }   \rho( {\cal C} )  w^{out}({\cal C})
    -  \sum_{{\cal C} } \sum_{{\cal C'} \ne {\cal C}  }  \sum_{r=1}^R  k_r({\cal C '} , {\cal C} )
\ln \left(  w_r({\cal C'} , {\cal C} )   \right)
\nonumber \\
&& = \sum_{{\cal C} } \sum_{{\cal C'} \ne {\cal C}  }    \sum_{r=1}^R
\left[    w_r({\cal C', \cal C}) \rho( {\cal C} )
    -   k_r ({\cal C '} , {\cal C} ) \ln \left(  w_r({\cal C'} , {\cal C} )   \right) \right]
\label{actionempirical}
\end{eqnarray}
where we have used Eq. \ref{wtoutr} in the second line
 in order to stress
the decomposition of the action onto the contributions of the reservoir dependent transition rates $w_r({\cal C'} , {\cal C} ) $.

In summary, 
the probability of an individual trajectory $[{\cal C}(t)]_{0 \leq t \leq T} $ of Eq. \ref{ptraject} decays exponentially in $T$
\begin{eqnarray}
  {\cal P} \left( [{\cal C}(t)]_{0 \leq t \leq T} \right) \opsimeq_{T \to +\infty} e^{- T A[ \rho(.), k_.(.,.) ]}
\label{ptrajectempi}
\end{eqnarray}
and the intensive action $A[ \rho(.), k_.(.,.) ] $
 only depends on the time-empirical density $\rho(\cal C)$ and the time-empirical jump densities
 $k_r(\cal C', \cal C)$ of the trajectory, that are thus the only relevant intensive observables of the trajectory.

This statement can be considered as the qualitative analog of Eq. \ref{pcano}
concerning the canonical ensemble, where the probability of a configuration 
 involves the extensive energy $E(\cal C) $ in the exponential,
so that the appropriate intensive  variable in the large $N$ limit
is the intensive energy $e=\frac{E(\cal C)}{N}$ of Eq. \ref{intensiveenergy}.

\subsection{ Intensive action $A[ \rho(.), j_.(.,.) , a_.(.,.)] $  in terms of the empirical currents and activities   }

Via the change of variables of Eq. \ref{jafromq},
the intensive action of Eq. \ref{actionempirical}
becomes in terms of the empirical currents $j_r({\cal C', \cal C})  $ and empirical activities $a_r({\cal C', \cal C}) $
\begin{eqnarray}
  A[ \rho(.), j_.(.,.) , a_.(.,.)]
  && = \sum_{{\cal C} }   \rho( {\cal C} )  w^{out}({\cal C})
 \nonumber \\&& 
- \sum_{{\cal C} } \sum_{{\cal C'} > {\cal C}  }    \sum_{r=1}^R
\left[  
       \frac{a_r({\cal C', \cal C})   }{2} \ln \left[  w_r({\cal C'} , {\cal C} ) w_r({\cal C} , {\cal C'} ) \right]
 +   \frac{j_r({\cal C', \cal C})   }{2} \ln \left(  \frac{ w_r({\cal C'} , {\cal C} )  }{w_r({\cal C} , {\cal C'} )}  \right)
\right]
\label{actionempiricalja}
\end{eqnarray}

In particular, one sees why the simplest choice of Eq. \ref{simple}
for the transitions rates is also the simplest choice from the point of view of this intensive action,
since the activities $a_r({\cal C', \cal C})   $ completely disappear, i.e. 
the intensive action $A^{simple}[ \rho(.), j_.(.,.)] $ only depends on $\rho( {\cal C} ) $ and $j_r({\cal C', \cal C})  $.

\subsection{ Entropy production in terms of the time-empirical heats $q_r$ of the trajectory  }

The probability 
of the backward trajectory ${\cal C}^{Backward}(t) \equiv {\cal C}(T-t) $
conditioned to start in configuration ${\cal C}_T$
involves the backward empirical observables of Eq. \ref{jabackward}
\begin{eqnarray}
  {\cal P} \left( [{\cal C}^{Backward}(t)]_{0 \leq t \leq T} \right) 
\opsimeq_{T \to +\infty}  e^{- T A[ \rho^{Backward}(.), j^{Backward}_.(.,.) , a^{Backward}_.(.,.)]  }
=  e^{- T A[ \rho(.), -  j_.(.,.) , a_.(.,.)]  }
\label{lnptrajectback}
\end{eqnarray}
As a consequence, the ratio between the probability of the initial trajectory $[{\cal C}(t)]_{0 \leq t \leq T} $
and the probability of the backward trajectory $[{\cal C}^{Backward}(t)]_{0 \leq t \leq T} $
only involves the contributions related to the empirical currents $j_r({\cal C', \cal C})  $
\begin{eqnarray}
 \frac{ {\cal P} \left( [{\cal C}(t)]_{0 \leq t \leq T} \right) }
{ {\cal P} \left( [{\cal C}^{Backward}(t)]_{0 \leq t \leq T} \right) }
=  e^{\displaystyle - T \left( A[ \rho(.),   j_.(.,.) , a_.(.,.)] -A[ \rho(.), -  j_.(.,.) , a_.(.,.)] \right) }
 =  e^{ \displaystyle T \sum_{{\cal C} } \sum_{{\cal C'} > {\cal C}  }    \sum_{r=1}^R   j_r({\cal C', \cal C})    \ln \left(  \frac{ w_r({\cal C'} , {\cal C} )  }{w_r({\cal C} , {\cal C'} )}  \right)
 }
\label{lnptrajectbackratio}
\end{eqnarray}
Since the generalized-detailed condition of Eq. \ref{detailedr}
 fixes the ratios of Eq. \ref{detailedratio}, Eq. \ref{lnptrajectbackratio}
reduces to
\begin{eqnarray}
 \frac{ {\cal P} \left( [{\cal C}(t)]_{0 \leq t \leq T} \right) }
{ {\cal P} \left( [{\cal C}^{Backward}(t)]_{0 \leq t \leq T} \right) }
 =  e^{ \displaystyle- T  
 \sum_{r=1}^R \beta_r 
\sum_{{\cal C} } \sum_{{\cal C'} > {\cal C}  }  
  j_r({\cal C', \cal C})   \left[ E({\cal C'})  -E({\cal C}) \right] 
 } 
= e^{ \displaystyle- T   \sum_{r=1}^R \beta_r q_r } 
\label{lnptrajectbackratioq}
\end{eqnarray}
and thus only involves the intensive time-empirical heats $q_r$ introduced in Eq \ref{qr} 
and the inverse-temperatures $\beta_r$ of the reservoirs.
The intensive entropy production $\sigma^{prod} \left( [{\cal C}(t)]_{0 \leq t \leq T} \right)$
associated to the long dynamical trajectory $ [{\cal C}(t)]_{0 \leq t \leq T} $,
that measures its irreversibility by the comparison with the probability of the backwards trajectory $ [{\cal C}^{Backward}(t)]_{0 \leq t \leq T} $
\begin{eqnarray}
\sigma^{prod} \left( [{\cal C}(t)]_{0 \leq t \leq T} \right) \equiv \frac{1}{T} \ln \left(  \frac{ {\cal P} \left( [{\cal C}(t)]_{0 \leq t \leq T} \right) }
{ {\cal P} \left( [{\cal C}^{Backward}(t)]_{0 \leq t \leq T} \right) } \right)
\label{sigmaproddef}
\end{eqnarray}
thus only involves the weighted sum of the empirical heats $q_r$ of the dynamical trajectory
\begin{eqnarray}
\sigma^{prod} \left( [{\cal C}(t)]_{0 \leq t \leq T} \right)
= -  \left( A[ \rho(.),   j_.(.,.) , a_.(.,.)] -A[ \rho(.), -  j_.(.,.) , a_.(.,.)] \right)
 =  -  \sum_{r=1}^R \beta_r q_r 
\label{sigmaprod}
\end{eqnarray}
with coefficients given by the 
inverse-temperatures $\beta_r$ of the reservoirs 
(while the non-weighted sum over the $q_r$ vanishes (Eq \ref{qrsum0})) :
this is the empirical version of the second principle of thermodynamics at the level of a very long dynamical trajectory
(see the review \cite{broeck} and references therein).

\subsection{ Relation between the action $ A[ \rho(.), k_.(.,.) ] $ and the Kolmogorov-Sinai entropy $h^{KS}$ }

In information theory and in stochastic thermodynamics, 
the 'surprise' ${\cal S} \left( [{\cal C}(t)]_{0 \leq t \leq T} \right) $
to observe as outcome the trajectory $[{\cal C}(t)]_{0 \leq t \leq T} $
when its probability is ${\cal P} \left( [{\cal C}(t)]_{0 \leq t \leq T} \right)  $ is defined as
minus the logarithm of this probability,
so that it coincides with the extensive action ${\cal A} \left( [{\cal C}(t)]_{0 \leq t \leq T} \right)  $ of Eq. \ref{lnptraject}
\begin{eqnarray}
 {\cal S} \left( [{\cal C}(t)]_{0 \leq t \leq T} \right) \equiv - \ln \left[  {\cal P} \left( [{\cal C}(t)]_{0 \leq t \leq T} \right) \right]
={\cal A} \left( [{\cal C}(t)]_{0 \leq t \leq T} \right)
\opsimeq_{T \to +\infty} T A[ \rho(.), k_.(.,.) ]
\label{surprise}
\end{eqnarray}
As a consequence,  the intensive action $ A[ \rho(.), k_.(.,.) ] $ of a dynamical trajectory
represents the 'intensive surprise' of the trajectory,
and depends only of the empirical observables $[ \rho(.), k_.(.,.) ]  $ of this trajectory.

The Shannon entropy $S^{dyn}_T$ associated to the trajectories probabilities ${\cal P} \left( [{\cal C}(t)]_{0 \leq t \leq T} \right)  $
then corresponds to the average of the surprise ${\cal S} \left( [{\cal C}(t)]_{0 \leq t \leq T} \right)  $ over the trajectories,
or to the average of the extensive action over the trajectories
\begin{eqnarray}
S^{dyn}_T && \equiv - \sum_{ [{\cal C}(t)]_{0 \leq t \leq T} } {\cal P} \left( [{\cal C}(t)]_{0 \leq t \leq T} \right) 
  \ln \left[  {\cal P} \left( [{\cal C}(t)]_{0 \leq t \leq T} \right) \right]
= <- \ln \left[  {\cal P} \left( [{\cal C}(t)]_{0 \leq t \leq T} \right) \right]  >
= < {\cal A} \left( [{\cal C}(t)]_{0 \leq t \leq T} \right) >
\nonumber \\
&& \opsimeq_{T \to +\infty} T  < A[ \rho(.), k_.(.,.) ] >
\label{hskse}
\end{eqnarray}
The Kolmogorov-Sinai entropy $h^{KS} $ defined as the coefficient of growth with respect to the time $T$ of Eq. \ref{hskse}
thus corresponds to the average of the intensive action $A[ \rho(.), k_.(.,.) ] $
\begin{eqnarray}
h^{KS} \equiv \lim_{T \to +\infty} \left( \frac{S^{dyn}_T }{T} \right)=  < A[ \rho(.), k_.(.,.) ] >
\label{hks}
\end{eqnarray}
over the dynamical trajectories, that will be translated into an average over the empirical variables in the next sections.


\section{ Entropy of the set of trajectories with given empirical observables  }

\label{sec_entropy}

\subsection{ Measure of the set of dynamical trajectories with given empirical observables }

Since all the individual dynamical trajectories $  [{\cal C}(t)]_{0 \leq t \leq T}  $
 that have the same empirical observables $\rho(\cal C)$ and 
 $k_r(\cal C', \cal C)$ have the same probability given by Eq. \ref{ptrajectempi},
 one can rewrite the normalization of Eq. \ref{normaformal} 
as a sum over these empirical observables
\begin{eqnarray}
&& 1  = \sum_{ \rho(.)  } \sum_{ k_.(.,.)  } 
\Omega_T[ \rho(.), k(.,.) ]
e^{- T A[ \rho(.), k(.,.) ] }
\label{normaempi}
\end{eqnarray}
where $\Omega_T[ \rho(.), k_.(.,.) ]$ represents the 
measure of the set of the dynamical trajectories associated to given values of these 
empirical observables.
So besides the empirical constraints ${\cal D} \left[ \rho(. ),  k_.(.,. ) \right] $ of Eq. \ref{dconstraint}, 
one expects an exponential growth with respect to $T$
\begin{eqnarray}
 \Omega_T[ \rho(.), k_.(.,.) ]  \opsimeq_{T \to +\infty}  {\cal D} \left[ \rho(. ),  k_.(.,. ) \right]
 e^{\displaystyle T S^{Boltzmann}[ \rho(.), k_.(.,.) ]  }
\label{omegat}
\end{eqnarray}
where $S^{Boltzmann}[ \rho(.), k_.(.,.) ]   $ represents the 
Boltzmann intensive entropy of the set of trajectories 
 with given intensive empirical observables.

This statement can be considered as the qualitative analog of Eq. \ref{entropymicrointensive}
concerning the microcanonical ensemble, where the 
number of configurations with a given intensive energy $e$ 
grows exponentially in $N$ with a coefficient given by 
the Boltzmann intensive entropy $s(e)$.

Then Eq \ref{normaempi} becomes
\begin{eqnarray}
 1  \opsimeq_{T \to +\infty} \sum_{ \rho(.)  } \sum_{ k_.(.,.)  }  {\cal D} \left( \rho(. ),  k_.(.,. ) \right)
e^{ T \left(S^{Boltzmann}[ \rho(.), k_.(.,.) ] -    A[ \rho(.), k_.(.,.) ] \right) }
\label{normaempit}
\end{eqnarray}

\subsection{ Analysis for typical values of the empirical observables }

The typical values of the empirical observables read
\begin{eqnarray}
   \rho^{typ}({\cal C}) && = \rho_{st}({\cal C})
\nonumber \\
k^{typ}_r({\cal C', \cal C}) && = w_r({\cal C', \cal C}) \rho_{st}({\cal C})
\label{typical}
\end{eqnarray}
 in terms of the normalized stationary solution $\rho_{st}({\cal C}) $ of
 the master Equation \ref{master}, that should satisfy for all ${\cal C}$
\begin{eqnarray}
0 && = \sum_{\cal C' \ne \cal C} \sum_{r=1}^R w_r({\cal C, \cal C'}) \rho_{st}({\cal C'}) 
- w^{out}({\cal C}) \rho_{st}({\cal C})
=  \sum_{\cal C' \ne \cal C} \left[ \sum_{r=1}^R w_r({\cal C, \cal C'}) \rho_{st}({\cal C'}) 
- \sum_{r=1}^R w_r({\cal C', \cal C}) \rho_{st}({\cal C}) \right]
\label{masterstatio}
\end{eqnarray}

For the typical values of Eq. \ref{typical}, 
the exponential behavior in $T$ of Eq. \ref{normaempit}
should exactly vanish,
i.e. the entropy $S^{Boltzmann}[ \rho^{typ}(.), k^{typ}_.(.,.) ]  $ should exactly compensate the action $A[ \rho^{typ}(.), k^{typ}_.(.,.) ]   $ of Eq. \ref{actionempirical}
\begin{eqnarray}
S^{Boltzmann}[ \rho^{typ}(.), k^{typ}_.(.,.) ] =    A[ \rho^{typ}(.), k^{typ}_.(.,.) ] 
=\sum_{{\cal C} } \sum_{{\cal C'} \ne {\cal C}  }    \sum_{r=1}^R
\left[    w_r({\cal C', \cal C}) \rho^{typ}( {\cal C} )
    -   k^{typ}_r ({\cal C '} , {\cal C} ) \ln \left(  w_r({\cal C'} , {\cal C} )   \right) \right]
\label{compensation}
\end{eqnarray}

\subsection{ Intensive entropy for generic values of the intensive empirical observables  }

To obtain the intensive entropy $S^{Boltzmann}[ \rho(.), k_.(.,.) ] $ 
 for generic values $\rho({\cal C} )$ and $k_r({\cal C', \cal C}) $ of the intensive empirical observables,
one just needs to introduce the effective transitions rates $w^{eff}_r({\cal C', \cal C}) $ that would make 
the empirical values typical for this effective modified dynamics, namely (Eq. \ref{typical})
\begin{eqnarray}
  w^{eff}_r({\cal C', \cal C}) = \frac{k_r({\cal C', \cal C})}{\rho({\cal C})}
\label{weff}
\end{eqnarray}
Then one may use Eq. \ref{compensation} with these effective rates to obtain
\begin{eqnarray}
S^{Boltzmann}[ \rho(.), k_.(.,.) ]
&& =\sum_{{\cal C} } \sum_{{\cal C'} \ne {\cal C}  }    \sum_{r=1}^R
\left[    w^{eff}_r({\cal C', \cal C}) \rho( {\cal C} )
    -   k_r ({\cal C '} , {\cal C} ) \ln \left(  w^{eff}_r({\cal C'} , {\cal C} )   \right) \right]
\nonumber \\
&& =\sum_{{\cal C} } \sum_{{\cal C'} \ne {\cal C}  }    \sum_{r=1}^R
\left[    k_r({\cal C', \cal C}) 
    -   k_r ({\cal C '} , {\cal C} ) \ln \left(  \frac{k_r({\cal C', \cal C})}{\rho({\cal C})}   \right) \right]
\label{entropyempi}
\end{eqnarray}
As explained after Eq \ref{omegat}, this intensive Boltzmann entropy $S^{Boltzmann}[ \rho(.), k_.(.,.) ]   $ 
of the set of dynamical trajectories that are characterized by the intensive empirical observables $ [ \rho(.), k_.(.,.) ]$
is the analog of the intensive microcanonical entropy $s(e)$ as a function of the intensive energy $e$ in
the microcanonical ensembles (Eq. \ref{entropymicrointensive}).
Here the explicit expression of Eq. \ref{entropyempi} was derived by the method of 'change of measure'
which is very common in the whole field of large deviations, since it allows to obtain
directly the result without any actual computation (i.e. one does not need 
to use combinatorial methods to count the appropriate configurations).

\subsection{ Alternative expression of the intensive entropy in terms of empirical activities and currents  }

Via the change of variables of Eq. \ref{jafromq},
the intensive entropy of Eq. \ref{entropyempi}
becomes in terms of the empirical currents and empirical activities
\begin{eqnarray}
 && 
S^{Boltzmann}[ \rho(.),  j_.(.,.) , a_.(.,.) ] 
 \nonumber \\  &&  
 = \sum_{{\cal C} } \sum_{{\cal C'} > {\cal C}  }    \sum_{r=1}^R
\left[  a_r({\cal C', \cal C}) 
    -   \frac{a_r({\cal C', \cal C})     }{2} 
\ln \left( \frac{  a^2_r({\cal C', \cal C}) -j_r^2({\cal C', \cal C})      }{4 \rho({\cal C'}) \rho({\cal C})}   \right) 
   -   \frac{j_r({\cal C', \cal C})    }{2} 
\ln \left( \frac{  a_r({\cal C', \cal C}) +j_r({\cal C', \cal C})      }{ a_r({\cal C', \cal C}) -j_r({\cal C', \cal C})   }  
\times \frac{\rho({\cal C'})}{\rho({\cal C})}  \right) 
\right] 
\label{entropyempija}
\end{eqnarray}

For the next section, it is interesting to note here the asymmetry property with respect to the empirical
currents $j_r({\cal C', \cal C})$, for any values of the empirical density $\rho({\cal C} )$ and activities $a_r({\cal C', \cal C}) $
(Eq \ref{jabackward})
\begin{eqnarray}
 && 
S^{Boltzmann}[ \rho(.),  j_.(.,.) , a_.(.,.) ] - S^{Boltzmann}[ \rho(.),  -j_.(.,.) , a_.(.,.) ] 
 = \sum_{{\cal C} } \sum_{{\cal C'} > {\cal C}  }    \sum_{r=1}^R
\frac{j_r({\cal C', \cal C})    }{2} 
\ln \left( \frac{\rho({\cal C})}{\rho({\cal C'})}  \right) 
\label{entropyasym}
\end{eqnarray}


\section{ Large deviations for the probability of intensive empirical densities }

\label{sec_2.5}

\subsection{ Joint probability of the empirical density $\rho({\cal C})$ and empirical jump densities $k_r({\cal C', \cal C})$ }

Eq \ref{normaempi}
means that the probability $ P_T \left[ \rho(. ),  k(.,. ) \right] $ of the empirical observables $\rho(. ) $ and $ k_.(.,. ) $
over the set of dynamical trajectories,
with the normalization
\begin{eqnarray}
&& 1  = \sum_{ \rho(.)  } \sum_{ k(.,.)  }  P_T \left[ \rho(. ),  k(.,. ) \right]
\label{normaprobaempi}
\end{eqnarray}
is given by
\begin{eqnarray}
&& P_T \left[ \rho(. ),  k_.(.,. ) \right]  =   
\Omega_T[ \rho(.), k(.,.) ]
e^{- T A[ \rho(.), k(.,.) ] }
\label{probaaempi}
\end{eqnarray}

Plugging Eq \ref{omegat}
 into Eq \ref{probaaempi}
yields that
the probability $P_T\left[ \rho(. ),  k_.(.,. ) \right]$ 
displays the large deviation form
\begin{eqnarray}
 P_T \left[ \rho(. ),  k_.(.,. ) \right] 
&& \opsimeq_{T \to +\infty}   {\cal D} \left[ \rho(. ),  k_.(.,. ) \right]
e^{- T I[ \rho(.),  k_.(. , .) ] }
\label{proba2.5}
\end{eqnarray}
where the rate function $ I[ \rho(.),  k_.(. , .) ] $
is simply given by the difference between the action
$A[ \rho(.), k_.(.,.) ] $ of Eq \ref{actionempirical} and 
the entropy $S^{Boltzmann}[ \rho(.), k_.(.,.) ]  $
of Eq \ref{entropyempi}
\begin{eqnarray}
 I[ \rho(.),  k_.(. , .) ] && = A[ \rho(.), k_.(.,.) ] - S^{Boltzmann}[ \rho(.), k_.(.,.) ] 
\nonumber \\
&& = 
\sum_{{\cal C} } \sum_{{\cal C'} \ne {\cal C}  }    \sum_{r=1}^R
\left[    w_r({\cal C', \cal C}) \rho( {\cal C} )
 - k_r({\cal C', \cal C}) 
    +
       k_r ({\cal C '} , {\cal C} ) \ln \left(  \frac{k_r({\cal C', \cal C})}{w_r({\cal C'} , {\cal C} )\rho({\cal C})}   \right) \right]
\label{rate2.5}
\end{eqnarray}
This formula is thus a very direct generalization in the presence of several reservoirs $r$
of the standard large deviation rate at level $2.5$ for Markov jump processes.
\cite{fortelle_thesis,fortelle_jump,maes_canonical,maes_onandbeyond,wynants_thesis,chetrite_formal,BFG1,BFG2}.
It vanishes only for the typical values of Eq. \ref{typical}
\begin{eqnarray}
 I[ \rho^{typ}(.),  k^{typ}_.(. , .) ] && = 0
\label{rate2.5zero}
\end{eqnarray}
while the positive values $ I[ \rho(.),  k_.(. , .) ] >0 $ for non-typical empirical observables $[ \rho(.),  k_.(. , .) ] \ne [ \rho^{typ}(.),  k^{typ}_.(. , .) ]  $
measure how rare it is to observe them for large time $T$.

Here the qualitative analogy with equilibrium is that the intensive free-energy of the canonical ensemble
contains an energetic contribution and an entropic contribution (Eq \ref{legendref}).

\subsection{ Joint probability of the empirical density, empirical activities and currents  }

Via the change of variables of Eq. \ref{jafromq},
the probability of Eq. \ref{proba2.5}
can be translated into the
joint probability of the empirical density, the empirical activities and the empirical currents 
\begin{eqnarray}
 P_T \left[ \rho(. ),  j_.(.,. ) ,  a_.(.,. )\right]
&& \opsimeq_{T \to +\infty}   {\cal D} \left[ \rho(. ),  j_.(.,. ) \right]
e^{- T I[ \rho(.),  j_.(. , .), a_.(. , .) ] }
\label{proba2.5ja}
\end{eqnarray}
where the constraints ${\cal D} \left[ \rho(. ),  j_.(.,. ) \right] $ are given
in terms of the currents in Eq \ref{dconstraintj},
while the rate function reads (Eq \ref{rate2.5})
\begin{eqnarray}
&& I[ \rho(.),  j(.,.) , a(.,.) ]  
 = \sum_{{\cal C} } \sum_{{\cal C'} > {\cal C}  }    \sum_{r=1}^R
\left[    w_r({\cal C', \cal C}) \rho( {\cal C} )+  w_r({\cal C, \cal C'}) \rho( {\cal C'} )
 - a_r({\cal C', \cal C}) 
\right] 
\label{rate2.5ja}
 \\
&& + \sum_{{\cal C} } \sum_{{\cal C'} > {\cal C}  }    \sum_{r=1}^R
\left[  
       \frac{a_r({\cal C', \cal C})     }{2} 
\ln \left( \frac{  a_r^2({\cal C', \cal C}) -j_r^2({\cal C', \cal C})      }
{4 w_r({\cal C} , {\cal C'} )\rho({\cal C'}) w_r({\cal C'} , {\cal C} )\rho({\cal C})}   \right) 
   +   \frac{j_r({\cal C', \cal C})    }{2} 
\ln \left( \frac{  a_r({\cal C', \cal C}) +j_r({\cal C', \cal C})      }{ a_r({\cal C', \cal C}) -j_r({\cal C', \cal C})   }  
\times \frac{w_r({\cal C} , {\cal C'} )\rho({\cal C'})}{ w_r({\cal C'} , {\cal C} )\rho({\cal C})}  \right) 
\right] 
\nonumber
\end{eqnarray}

An important property of this form of the rate function
is that
 the ratio of the probabilities to observe the empirical currents $j_r({\cal C', \cal C})$ 
and to observe the opposite empirical currents $(-j_r({\cal C', \cal C}))$
that would characterize the Backwards trajectories (Eq \ref{jabackward})
simplifies into a linear term with respect to the currents in the exponential 
for any empirical density $\rho({\cal C} )$ and activities $a_r({\cal C', \cal C}) $
\begin{eqnarray}
\frac{  P_T \left[ \rho(. ),  j_.(.,. )  ,  a_.(.,. )\right] }{  P_T \left[ \rho(. ),  -j_.(.,. )  ,  a_.(.,. )\right] }
&& \oppropto_{T \to +\infty} e^{ \displaystyle 
- T \left[ I[ \rho(.),  j_.(. , .)  ,  a_.(.,. )] - I[ \rho(.),  -j_.(. , .)  ,  a_.(.,. )]\right]} 
 \nonumber \\ && 
= e^{ \displaystyle T  \sum_{{\cal C} } \sum_{{\cal C'} > {\cal C}  }     \sum_{r=1}^R 
 j_r({\cal C', \cal C})    
\ln \left( \frac{ w_r({\cal C'} , {\cal C} )\rho({\cal C})     }{  w_r({\cal C} , {\cal C'} )\rho({\cal C'})}  \right)  
} 
\label{symj}
\end{eqnarray}
Again this is a very direct generalization in the presence of several reservoirs $r$
of the asymmetry with respect to the empirical currents \cite{maes_canonical,c_interactions} :
one recognizes the contribution of the action asymmetry (Eq. \ref{sigmaprod})
and the contribution of the entropy asymmetry (Eq. \ref{entropyasym}) discussed previously.

\subsection{ Joint probability of the empirical density and the empirical currents (without the empirical activities) }

From Eq \ref{proba2.5ja}, one can obtain
the joint probability of the empirical density and the empirical currents
 by integrating over the empirical activities that do not appear in the constraints
\begin{eqnarray}
 P_T \left[ \rho(. ),  j_.(.,. ) \right]  = \sum_{a_.(.,. ) } P_T \left[ \rho(. ),  j_.(.,. ) ,  a_.(.,. )\right]
&& \opsimeq_{T \to +\infty}   {\cal D} \left[ \rho(. ),  j_.(.,. ) \right]
\sum_{a_.(.,. ) } e^{- T I[ \rho(.),  j_.(. , .), a_.(. , .) ] }
\label{proba2.5j}
\end{eqnarray}
As a consequence, one just needs to optimize the rate function $I[ \rho(.),  j_.(.,.) , a_.(.,.) ] $
over the activities $ a_r({\cal C', \cal C})$
\begin{eqnarray}
 && 
0 = \frac{ \partial I[ \rho(.),  j_.(.,.) , a_.(.,.) ] }{ \partial a_r({\cal C', \cal C}) }
 = 
       \frac{1     }{2} 
\ln \left( \frac{  a_r^2({\cal C', \cal C}) -j_r^2({\cal C', \cal C})      }{4 w_r({\cal C, \cal C'})\rho({\cal C'}) w_r({\cal C', \cal C})\rho({\cal C})}   \right) 
\label{deri2.5}
\end{eqnarray}
Plugging the optimal solution
\begin{eqnarray}
 a_r^{opt} ({\cal C', \cal C})= \sqrt{ j_r^2({\cal C', \cal C}) + 4 w_r({\cal C, \cal C'})\rho({\cal C'}) w_r({\cal C', \cal C})\rho({\cal C})}  
\label{aopt2.5}
\end{eqnarray}
into the rate function yields the optimal value
\begin{eqnarray}
    I^{opt}[ \rho(.),  j_.(.,.) ]  && =  I[ \rho(.),  j_.(.,.) , a_.^{opt}(.,.) ]  
 \nonumber \\  &&  
= \sum_{{\cal C} } \sum_{{\cal C'} > {\cal C}  }     \sum_{r=1}^R
\left[    w_r({\cal C', \cal C}) \rho( {\cal C} )+  w_r({\cal C, \cal C'}) \rho( {\cal C'} )
 - \sqrt{ j^2({\cal C', \cal C}) + 4 w_r({\cal C, \cal C'})\rho({\cal C'}) w_r({\cal C', \cal C})\rho({\cal C})}  
\right]
\nonumber \\
&& + \sum_{{\cal C} } \sum_{{\cal C'} > {\cal C}  }     \sum_{r=1}^R
\left[    j_r({\cal C', \cal C})    
\ln \left( \frac{ \sqrt{ j_r^2({\cal C', \cal C}) + 4 w_r({\cal C, \cal C'})\rho({\cal C'}) w_r({\cal C', \cal C})\rho({\cal C})}  +j_r({\cal C', \cal C})      }{ 2 w_r({\cal C'} , {\cal C} )\rho({\cal C})}  \right) 
\right] 
\label{i2.5rhoj}
\end{eqnarray}
that governs the large deviation form of Eq. \ref{proba2.5j}
\begin{eqnarray}
 P_T \left[ \rho(. ),  j_.(.,. ) \right]  \opsimeq_{T \to +\infty}   {\cal D} \left[\rho(. ),  j_.(.,. ) \right]
e^{- T I^{opt}[ \rho(.),  j_.(. , .) ] }
\label{proba2.5jld}
\end{eqnarray}
Again Eq \ref{i2.5rhoj} is a very direct generalization in the presence of several reservoirs $r$
of the known formula for Markov jump processes \cite{maes_canonical,c_ring,c_interactions}.

Eq. \ref{i2.5rhoj} yields the property directly inherited from Eq. \ref{symj}
\begin{eqnarray}
\frac{  P_T \left[ \rho(. ),  j_.(.,. ) \right] }{  P_T \left[ \rho(. ),  -j_.(.,. ) \right] }
&& \oppropto_{T \to +\infty} e^{ \displaystyle 
- T \left[ I^{opt}[ \rho(.),  j_.(. , .) ] - I^{opt}[ \rho(.),  -j_.(. , .) ]\right]} 
 \nonumber \\ && 
= e^{ \displaystyle T  \sum_{{\cal C} } \sum_{{\cal C'} > {\cal C}  }     \sum_{r=1}^R 
 j_r({\cal C', \cal C})    
\ln \left( \frac{ w_r({\cal C'} , {\cal C} )\rho({\cal C})     }{  w_r({\cal C} , {\cal C'} )\rho({\cal C'})}  \right)  
} 
\label{symjbis}
\end{eqnarray}


\section{ Statistics of the other intensive time-empirical observables }

\label{sec_additive}

In this section, we explain for our present framework with several thermal reservoirs
how the statistics of any intensive empirical time-additive observable
of the dynamical trajectory 
can be reconstructed from its decomposition
 in terms of the empirical densities and of the empirical jump densities studied in the previous sections.
We also mention the link with the generating function method
based on deformed Markov operators, that have attracted a lot of interest recently
in various models
 \cite{derrida-lecture,lecomte_chaotic,lecomte_thermo,lecomte_formalism,
lecomte_glass,kristina1,kristina2,chetrite_canonical,
chetrite_conditioned,lazarescu_companion,lazarescu_generic,touchette_langevin,derrida-conditioned,bertin-conditioned}.

\subsection{ Decomposition onto the time empirical density and time-empirical jump densities }

For the trajectory $[{\cal C}(t)]_{0 \leq t \leq T} $ of Eq. \ref{traject},
any intensive time-empirical observable $b \left( [{\cal C}(t)]_{0 \leq t \leq T} \right)  $ of the form
(where $f({\cal C})$ and $g_r({\cal C'},{\cal C} )$ for $ {\cal C'} \ne {\cal C}$ are two arbitrary functions)
\begin{eqnarray}
  b \left( [{\cal C}(t)]_{0 \leq t \leq T} \right) 
\equiv \frac{1}{T}   \int_0^T dt f({\cal C}(t))     
+ \frac{1}{T} \sum_{k=1}^M g_{r_k}({\cal C}(t_k^+) , {\cal C}(t_k^-) )  
\label{btraject}
\end{eqnarray}
only depends on the time-empirical density $\rho({\cal C})$ 
and the time-empirical jump densities
$k_r({\cal C '} , {\cal C} )$ introduced in subsection \ref{sub_empi}
\begin{eqnarray}
  b \left( [{\cal C}(t)]_{0 \leq t \leq T} \right) 
&& =  \sum_{{\cal C} }   \rho( {\cal C} )  f({\cal C})
    +  \sum_{{\cal C} } \sum_{{\cal C'} \ne {\cal C}  }  \sum_{r=1}^R  k_r({\cal C '} , {\cal C} )
g_r({\cal C'} , {\cal C} )  
\equiv b\left[ \rho(.),k_.(.,.)  \right]
\label{bempirical}
\end{eqnarray}

\subsection{ Large deviation form of the probability distribution $p_T(b) $ of $b$  }

Its probability distribution over trajectories $[{\cal C}(t)]_{0 \leq t \leq T} $
with the measure of Eq. \ref{normaformal}
\begin{eqnarray}
p_T(b)  \equiv \sum_{ [{\cal C}(t)]_{0 \leq t \leq T} } {\cal P} \left( [{\cal C}(t)]_{0 \leq t \leq T} \right) \ \delta\left( b- b \left( [{\cal C}(t)]_{0 \leq t \leq T} \right)  \right)
\label{pbtraj}
\end{eqnarray}
can be thus rewritten in terms of the probability distribution of 
the intensive empirical observables with the measure of Eq. \ref{normaprobaempi}
with the large deviation form of Eq. \ref{proba2.5}
\begin{eqnarray}
 p_T(b) 
&& = \sum_{ \rho(.)  } \sum_{ k(.,.)  }  {\cal P}_T \left[ \rho(. ),  k(.,. ) \right]
\ \delta\left( b- b \left[  \rho(.),k_.(.,.)       \right]  \right)
\nonumber \\
&&
\opsimeq_{T \to +\infty}
 \sum_{ \rho(.)  } \sum_{ k(.,.)  } 
  {\cal D} \left[ \rho(. ),  k_.(.,. ) \right]
\ \delta\left( b- b \left[  \rho(.),k_.(.,.)       \right]  \right) \ 
e^{- T I[ \rho(.),  k_.(. , .) ] }
\label{pbempi}
\end{eqnarray}
where the constraint $ \delta\left( b- b \left[  \rho(.),k_.(.,.)       \right]  \right)$
appears as a supplementary constraint with respect to the constitutive ones ${\cal D} \left( \rho(. ),  k_.(.,. ) \right) $
of Eq. \ref{dconstraint}.
As a consequence, the probability distribution of $b$ will follow the large deviation form
\begin{eqnarray}
 p_T(b) 
\opsimeq_{T \to +\infty}   e^{- T L( b ) }
\label{pblargedev}
\end{eqnarray}
where the rate function $L( b ) $ corresponds to the minimum of the rate functional $I[ \rho(.),  k_.(. , .) ] $ over $\rho(.)$ and $k_.(.,.) $ 
in the presence of the constraints $ \delta\left( b- b \left[  \rho(.),k_.(.,.)       \right]  \right) $ and ${\cal D} \left[ \rho(. ),  k_.(.,. ) \right] $ : this operation is generically called 'contraction' in the large deviation language.

\subsection{ Scaled cumulant generating function of $b$ via the appropriate deformed Markov operator  }

The large deviation form of Eq. \ref{pblargedev}
translates into the large deviation form for generating function of $b$
\begin{eqnarray}
\int db p_T(b) e^{T \nu b } 
&& \opsimeq_{T \to +\infty}  \int db e^{T (\nu b - L(b) ) } \opsimeq_{T \to +\infty} e^{T \phi(\nu) }
\label{genepbempi}
\end{eqnarray}
where the saddle-point evaluation of the integral over $b$
yields that the scaled cumulant generating function $\phi(\nu)$
is the Legendre transform of the rate function $L(b)$
\begin{eqnarray}
\phi(\nu)  && = \sup_{b} \left( \nu b - L(b) \right)
\label{legendre}
\end{eqnarray}

The generating function reads in terms of the probability $ {\cal P}_T \left[ \rho(. ),  k(.,. ) \right] $
of the empirical densities
\begin{eqnarray}
\int db p_T(b) e^{\nu (T  b) } 
&& = \sum_{ \rho(.)  } \sum_{ k(.,.)  }  {\cal P}_T \left( \rho(. ),  k(.,. ) \right)
e^{ T \nu b \left[  \rho(.),k_.(.,.)       \right]  }
\nonumber \\
&&
\opsimeq_{T \to +\infty}
 \sum_{ \rho(.)  } \sum_{ k(.,.)  } 
  {\cal D} \left[ \rho(. ),  k_.(.,. ) \right]
e^{ T \left( \nu b \left[  \rho(.),k_.(.,.)       \right]-  I[ \rho(.),  k_.(. , .) ] \right)  }
\label{genpbempi}
\end{eqnarray}

Using the explicit expressions of Eq. \ref{bempirical}
and Eq. \ref{rate2.5},
the factor of $T$ in the exponential reads 
\begin{eqnarray}
 && \nu b \left[  \rho(.),k_.(.,.)       \right]-  I[ \rho(.),  k_.(. , .) ] 
 = \nu \sum_{{\cal C} }   \rho( {\cal C} )  f({\cal C})
    + \nu  \sum_{{\cal C} } \sum_{{\cal C'} \ne {\cal C}  }  \sum_{r=1}^R  k_r({\cal C '} , {\cal C} )
g_r({\cal C'} , {\cal C} )  
\nonumber \\
&& -\sum_{{\cal C} } \sum_{{\cal C'} \ne {\cal C}  }    \sum_{r=1}^R
\left[    w_r({\cal C', \cal C}) \rho( {\cal C} )
 - k_r({\cal C', \cal C}) 
    +
       k_r ({\cal C '} , {\cal C} ) \ln \left(  \frac{k_r({\cal C', \cal C})}{w_r({\cal C'} , {\cal C} )\rho({\cal C})}   \right) \right]
\nonumber \\
 && =  -\sum_{{\cal C} } 
\left[ \sum_{{\cal C'} \ne {\cal C}  }    \sum_{r=1}^R
    w_r({\cal C', \cal C})  -  \nu   f({\cal C})    \right] \rho( {\cal C} ) 
-\sum_{{\cal C} } \sum_{{\cal C'} \ne {\cal C}  }    \sum_{r=1}^R
k_r({\cal C', \cal C}) 
\left[    - 1   +    \ln \left(  \frac{k_r({\cal C', \cal C})}{e^{\nu g_r({\cal C'} , {\cal C} )  }w_r({\cal C'} , {\cal C} )\rho({\cal C})}   \right) \right]
\label{kbmi}
\end{eqnarray}

This type of result for Markov jump processes can be re-interpreted as
the effect of some $\nu-$dependent deformed Markov operator
for a vector $ \Psi^{(\nu)}_t(\cal C)$ (see the detailed proof in the section 4.2
of Reference \cite{chetrite_formal}) 
\begin{eqnarray}
\partial_t \Psi^{(\nu)}_t(\cal C) && = \sum_{\cal C' \ne \cal C}  
\sum_{r=1}^R  w^{\nu}_r ({\cal C', \cal C}) 
 \Psi^{(\nu)}_t({\cal C'}) - w^{(\nu)out}({\cal C}) \Psi^{(\nu)}_t({\cal C})
\label{mastertilted}
\end{eqnarray}
where the reservoir-dependent transition rates for $\cal C' \ne \cal C $
are tilted according to
\begin{eqnarray}
 w^{\nu}_r({\cal C', \cal C}) \equiv  e^{\nu g_r({\cal C'} , {\cal C} )  }  w_r ({\cal C', \cal C}) 
\label{tiltedoff}
\end{eqnarray}
while the diagonal terms are changed into
\begin{eqnarray}
w^{(\nu) out}( {\cal C}) \equiv  \sum_{{\cal C'} \ne  \cal C }  \sum_{r=1}^R  w_r ({\cal C', \cal C})
- \nu f({\cal C})
\label{tilteddiago}
\end{eqnarray}
Again, this is a very direct generalization in the presence of several reservoirs $r$
of the known result for Markov jump dynamics \cite{chetrite_formal,chetrite_conditioned,c_ring}.
Then the scaled cumulant generating function $\phi(\nu)$ of Eq. \ref{genepbempi}
corresponds to the highest eigenvalue of thee deformed Markov operator $w^{(\nu)}$ of Eqs \ref{tiltedoff} and \ref{tilteddiago},
while the corresponding left and right eigenvectors are essential to understand the properties of the 'conditioned' dynamics
(see \cite{chetrite_canonical,chetrite_conditioned,derrida-conditioned,bertin-conditioned,chetrite_HDR} and references therein).



\section{ Conclusion }

\label{sec_conclusion}

For a system in contact with several reservoirs $r$ at different inverse-temperatures $\beta_r$,
we have described how the Markov jump dynamics with the generalized detailed balance condition
could be analyzed via a statistical physics  approach of dynamical trajectories $[{\cal C}(t)]_{0 \leq t \leq T}   $ over a long time interval $T \to + \infty$. 
At each step, we have stressed the similarities and differences with the statistical physics theory of equilibrium.
The main points can be summarized as follows :

(i) the relevant intensive variables are the time-empirical density $\rho(\cal C)$ and  
the time-empirical jump densities $k_r ({\cal C', \cal C})  $ from configuration
${\cal C}  $ to configuration ${\cal C '}  $ produced by the reservoir $r$ (that furnishes or absorbs the energy difference).
The probability of an individual trajectory ${\cal P} \left( [{\cal C}(t)]_{0 \leq t \leq T} \right) \sim e^{- T A[ \rho(.), k_.(.,.) ]}$
involves the intensive action $A[ \rho(.), k_.(.,.) ] $ that only depends on these empirical observables.
The analogy with equilibrium is that, in the canonical ensemble, the intensive energy $e=\frac{E(\cal C)}{N}$
of a configuration is the only relevant intensive variable that determines the probability of a configuration in the thermodynamic limit $N \to +\infty$.

(ii) the measure $\Omega_T[ \rho(.), k_.(.,.) ]$ of the set of dynamical trajectories $[{\cal C}(t)]_{0 \leq t \leq T}  $
that have the same empirical variables grows exponentially 
$\Omega_T[ \rho(.), k_.(.,.) ]  \sim e^{\displaystyle T S^{Boltzmann}[ \rho(.), k_.(.,.) ]  }$ with respect to $T$,
where $  S^{Boltzmann}[ \rho(.), k_.(.,.) ]  $ represents the intensive Boltzmann entropy.
The analogy with equilibrium is that, in the microcanonical ensemble, the 
number of configurations with a given intensive energy $e$ 
grows exponentially in $N$ with a coefficient given by the Boltzmann intensive entropy $s(e)$.

(iii) the measure over dynamical trajectories ${\cal P} \left( [{\cal C}(t)]_{0 \leq t \leq T} \right) $
can be thus replaced by a measure $P_T[ \rho(.), k_.(.,.) ]$ over the empirical observables with their constraints :
the large deviation form for large $T$ involves the rate function $ I[ \rho(.),  k_.(. , .) ]  = A[ \rho(.), k_.(.,.) ] - S^{Boltzmann}[ \rho(.), k_.(.,.) ]  $ that involves the action of (i) and the entropy of (ii).
The analogy with equilibrium is that the intensive free-energy contains an energetic contribution and an entropic contribution.

(iv) This measure $ P_T[ \rho(.), k_.(.,.) ]$ over the relevant empirical variables
allows to study the statistics of any intensive time-additive observable $b$ of the dynamical trajectory,
in particular those that contain information on the exchanges with the various reservoirs.

(v) The behavior of various observables upon time-reversal has been analyzed via
the replacement of the time-empirical jump densities $k_r ({\cal C', \cal C})  $
by the time-empirical activities $a_r ({\cal C', \cal C})$  and currents $j_r ({\cal C', \cal C})$.

This non-equilibrium framework for a system in contact with several reservoirs $r$
has been formulated at the same level of generality of the statistical physics theory of Equilibrium,
where the system can be in a certain number of configurations $\cal C  $ of energies $E(\cal C)$.
For applications to specific models, besides this space of configurations with the energy function $ E(\cal C)$,
one needs to choose the transition rates $w_r({\cal C',\cal C})$, i.e. what dynamical elementary moves ${\cal C'} \leftarrow {\cal C} $ are allowed between configurations
(for instance single-spin-flip dynamics in spin models) and how the various reservoirs $r$ are involved in these possible transitions, depending on the real-space structure of the model under study.

As a final remark, let us note that here we have focused on a system that could only exchange energy 
with the various thermal reservoirs,
but a closely-related framework for systems exchanging particles with reservoirs
in terms of empirical densities and empirical flows
can be found in \cite{c_open} for independent particles and in \cite{c_interactions} for interacting particles,
with possibly time-dependent rates, in order to take into account the possibility 
of external driving with some protocol $\lambda(t)$, in particular for the case of periodic driving,
which has attracted a lot of interest recently from the point of view of large deviations \cite{c_open,singh,bertini,barato17,barato18}.


\appendix

\section{ Short reminder on the statistical physics theory of equilibrium }

\label{app_eq}

In this Appendix, we recall some basic ideas of the statistical physics of equilibrium
in order to make more explicit the analogy with some equations of the text concerning non-equilibrium.

\subsection{ Enumeration of the configurations ${\cal C}$ and of their energies $E({\cal C})$ }

\label{sec_config}

The statistical physics theory of equilibrium is formulated at a very general level
where a system is described by a set of configurations ${\cal C}$ characterized by their energies $E({\cal C})$.
To simplify the discussion, let us focus on the
 standard example of the the Ising model of $N$ classical spins $S_j = \pm 1$ in dimension $d$
with nearest-neighbor coupling $J$. The ${\cal N}=2^N$ configurations ${\cal C}=(S_1,...,S_N)$ 
can be decomposed with respect to the discrete extensive energy $E$
\begin{eqnarray}
{\cal N} =2^N = \sum_E {\cal N}(E) 
\label{sumne}
\end{eqnarray}
where
\begin{eqnarray}
 {\cal N}(E) \equiv  \sum_{\cal C} \delta_{E,E({\cal C})} 
\label{defne}
\end{eqnarray}
counts the number of configurations ${\cal C}$
 that have the same energy $E({\cal C})=E$.

\subsection{ Microcanonical ensemble at fixed energy $E$ (for an isolated system)  }

In the microcanonical ensemble associated to the fixed energy $E$,
the $({\cal N} - {\cal N}(E)   )$ configurations having a different energy $E({\cal C} )\ne E$ are not possible,
while the ${\cal N}(E)$ configurations having this energy $E({\cal C} )= E$ are equiprobable
\begin{eqnarray}
P^{micro}_{E}({\cal C}) \equiv \frac{\delta_{E,E({\cal C})}}{{\cal N}(E)} 
\label{pmicroe}
\end{eqnarray}
The corresponding Shannon entropy of this microcanonical probability
coincides with the Boltzmann entropy 
\begin{eqnarray}
S^{micro}_E \equiv  - \sum_{\cal C} P^{micro}_{E}({\cal C}) \ln \left(P^{micro}_{E}({\cal C}) \right) = \ln {\cal N}(E) \equiv S^{Boltzmann}_E 
\label{entropymicro}
\end{eqnarray}

For a large number $N$ of spins, both the energy $E$ and the entropy $S^{micro}_E $ are extensive in $N$.
 It is then useful to introduce the intensive energy 
\begin{eqnarray}
e && \equiv \frac{E}{N}
\label{intensiveenergy}
\end{eqnarray}
as well as the intensive entropy $s^{micro}(e)$ as a function of the intensive energy $e$
\begin{eqnarray}
s^{micro}(e) && \equiv \frac{ S^{micro}_{(E=Ne)} }{N}
\label{seintensive}
\end{eqnarray}
that governs the exponential growth in $N$ of the number of configurations with a given intensive energy $e$ 
\begin{eqnarray}
 {\cal N}(e) \opsimeq_{N \to +\infty} e^{ N s^{micro}(e) }
\label{entropymicrointensive}
\end{eqnarray}

The inverse temperature associated to this microcanonical ensemble reads
\begin{eqnarray}
\beta^{micro} = \frac{d s^{micro}(e) }{de} 
\label{tempmicro}
\end{eqnarray}

\subsection{ Canonical ensemble at inverse temperature $\beta$ (for a system in contact with a thermal reservoir)}

The canonical ensemble at inverse temperature $\beta$ corresponds to the probability distribution
\begin{eqnarray}
P^{cano}_{\beta}({\cal C}) \equiv \frac{ e^{- \beta E({\cal C}) } }{ Z_{\beta}} 
\label{pcano}
\end{eqnarray}
where the canonical partition function $ Z_{\beta}$ that ensures the normalization allows to define the free-energy $F_{\beta} $
via
\begin{eqnarray}
 e^{- \beta F_{\beta} } \equiv Z_{\beta}= \sum_{\cal C} e^{- \beta E({\cal C}) } = \sum_E {\cal N}(E) e^{- \beta E} = \sum_E  e^{S^{micro}_E- \beta E} 
\label{zbeta}
\end{eqnarray}

The corresponding Shannon entropy reads
\begin{eqnarray}
S^{cano}_{\beta} = - \sum_{\cal C} P^{cano}_{\beta}({\cal C}) \ln \left( P^{cano}_{\beta}({\cal C}) \right) 
= \beta \left( F_{\beta}-U_{\beta}  \right)
\label{entropycano}
\end{eqnarray}
in terms of the free-energy $F(\beta) $ of Eq. \ref{zbeta}
and of the internal energy 
\begin{eqnarray}
U_{\beta} =  \sum_{\cal C} E({\cal C}) P^{cano}_{\beta}({\cal C}) 
\label{energycano}
\end{eqnarray}

For small $N$, the canonical ensemble is of course completely different from the microcanonical ensemble. 
However for large $N$, the extensivity of the energy (Eq. \ref{intensiveenergy}) and of the microcanonical
entropy  (Eq. \ref{seintensive})
leads to the saddle-point evaluation of the partition function of Eq. \ref{zbeta}
\begin{eqnarray}
e^{- \beta F_{\beta} } \equiv Z_{\beta}  \simeq N \int de   \ \ e^{ N \left( s^{micro}(e)- \beta e \right) } 
\simeq e^{- \beta N f(\beta) }
\label{zbetasaddle}
\end{eqnarray}
where the intensive free-energy
\begin{eqnarray}
f(\beta) && \equiv \frac{ F(\beta) }{N}
\label{freeintensive}
\end{eqnarray}
is given by the saddle-point value of Eq. \ref{zbetasaddle}
\begin{eqnarray}
0 && = \partial_e \left( s^{micro}(e)- \beta e \right) = \beta^{micro}(e) - \beta
\nonumber \\
- \beta  f(\beta) && = s^{micro}(e)- \beta e
\label{legendref}
\end{eqnarray}
so that the canonical ensemble at inverse temperature $\beta$ is actually dominated in the thermodynamic limit
by
the configurations of the microcanonical ensemble at intensive energy $e$ satisfying $\beta^{micro}(e) = \beta $.



\begin{thebibliography}{99}

\bibitem{oono}
Y. Oono,
Progress of Theoretical Physics Supplement 99, 165 (1989).

\bibitem{ellis}
R.S. Ellis, Physica D 133, 106 (1999).

\bibitem{review_touchette}
H. Touchette, Phys. Rep. 478, 1 (2009).


\bibitem{fortelle_thesis}
A. de La Fortelle, PhD (2000)
"Contributions to the theory of large deviations and applications" INRIA Rocquencourt.

\bibitem{fortelle_chain}
G. Fayolle and A. de La Fortelle,
Problems of Information Transmission 38, 354 (2002).


\bibitem{c_largedevdisorder}
C. Monthus, Eur. Phys. J. B 92, 149 (2019) in the
topical issue " Recent Advances in the Theory of Disordered Systems"
edited by F. Igloi and H. Rieger.



\bibitem{fortelle_jump}
A. de La Fortelle, 
Problems of Information Transmission 37 , 120 (2001).



\bibitem{maes_canonical}
C. Maes and K. Netocny, Europhys. Lett. 82, 30003 (2008)

\bibitem{maes_onandbeyond}
C. Maes, K. Netocny and B. Wynants, Markov Proc. Rel. Fields. 14, 445 (2008).


\bibitem{chetrite_formal}
A. C. Barato and R. Ch\'etrite, J. Stat. Phys. 160, 1154 (2015).

\bibitem{BFG1}
L. Bertini, A. Faggionato and D. Gabrielli, 
Ann. Inst. Henri Poincare Prob. and Stat. 51, 867 (2015).

\bibitem{BFG2}
L. Bertini, A. Faggionato and D. Gabrielli, 
Stoch. Process. Appli. 125, 2786 (2015).


\bibitem{wynants_thesis}
B. Wynants, arXiv:1011.4210, PhD Thesis (2010), "Structures of Nonequilibrium Fluctuations", Catholic University of Leuven.



\bibitem{maes_diffusion}
C. Maes, K. Netocny and B.  Wynants
Physica A 387, 2675 (2008).


\bibitem{engel}
J. Hoppenau, D. Nickelsen and A. Engel,
 New J. Phys. 18 083010 (2016).



\bibitem{derrida-lecture}
B. Derrida, JSTAT P07023 (2007).



\bibitem{lecomte_chaotic}
V. Lecomte, C. Appert-Rolland and F. van Wijland,
Phys. Rev. Lett. 95 010601 (2005).

\bibitem{lecomte_thermo}
V. Lecomte, C. Appert-Rolland and F. van Wijland,
J. Stat. Phys. 127 51-106 (2007).

\bibitem{lecomte_formalism}
V. Lecomte, C. Appert-Rolland and F. van Wijland,
Comptes Rendus Physique 8, 609 (2007).

\bibitem{lecomte_glass}
J.P. Garrahan, R.L. Jack, V. Lecomte, E. Pitard, K. van Duijvendijk, F. van Wijland,
Phys. Rev. Lett. 98, 195702 (2007).

\bibitem{kristina1}
J.P. Garrahan, R.L. Jack, V. Lecomte, E. Pitard, K. van Duijvendijk and F. van Wijland, 
J. Phys. A 42, 075007 (2009).

\bibitem{kristina2}
K. van Duijvendijk, R.L. Jack and F. van Wijland, 
Phys. Rev. E 81, 011110 (2010).

\bibitem{chetrite_canonical}
R. Ch\'etrite and H. Touchette,
Phys. Rev. Lett. 111, 120601 (2013).

\bibitem{chetrite_conditioned}
R. Ch\'etrite and H. Touchette
 Ann. Henri Poincare 16, 2005 (2015).


\bibitem{lazarescu_companion}
A. Lazarescu, J. Phys. A: Math. Theor. 48 503001 (2015).

\bibitem{lazarescu_generic}
A. Lazarescu, J. Phys. A: Math. Theor. 50 254004 (2017).



\bibitem{touchette_langevin}
H. Touchette, Physica A 504, 5 (2018).

\bibitem{derrida-conditioned}
B. Derrida and T. Sadhu, 
Journal of Statistical Physics 176, 773 (2019)
and 
Journal of Statistical Physics 177, 151 (2019).



\bibitem{bertin-conditioned}
N. Tizon-Escamilla, V. Lecomte and E. Bertin, 	J. Stat. Mech. (2019) 013201.



\bibitem{harris_Schu}
R J Harris and G M Sch\"utz,
J. Stat. Mech.  P07020 (2007).

\bibitem{searles}
E.M. Sevick, R. Prabhakar, S. R. Williams, D. J. Searles,
Ann. Rev. of Phys. Chem.  Vol 59, 603 (2008). 

\bibitem{harris}
H. Touchette and R.J. Harris, chapter "Large deviation approach to nonequilibrium systems"
of the book "Nonequilibrium Statistical Physics of Small Systems: Fluctuation Relations and Beyond", Wiley 2013.

\bibitem{mft}
L. Bertini, A. De Sole, D. Gabrielli, G. Jona-Lasinio, and C. Landim
Rev. Mod. Phys. 87, 593 (2015).


\bibitem{vivien_thesis}
V. Lecomte, PhD Thesis (2007)
"Thermodynamique des histoires et fluctuations hors d'\'equilibre"
Universit\'e Paris 7.


\bibitem{chetrite_thesis}
R. Ch\'etrite, PhD Thesis 2008 
"Grandes d\'eviations et relations de fluctuation dans certains mod\`eles de syst\`emes
hors d'\'equilibre"  ENS Lyon

\bibitem{chetrite_HDR}
R. Ch\'etrite, HDR Thesis (2018)
"P\'er\'egrinations sur les ph\'enom\`enes al\'eatoires dans la nature",
 Laboratoire J.A. Dieudonn\'e, Universit\' e de Nice.




\bibitem{broeck}
C. Van Den Broeck, " Stochastic thermodynamics: A brief introduction"
in the book "Physics of Complex Colloids" Editors C. Bechinger, F. Sciortino, P. Ziherl (2013).

\bibitem{rates_simple1977}
E. D. Siggia, Phys. Rev. B 16, 2319 (1977).


\bibitem{rates_simple2013}
C. Monthus and T. Garel, J. Stat. Mech. (2013) P02037

\bibitem{rates_simple2016}
C. Monthus, J. Stat. Mech. (2016) 043302

\bibitem{glauber}
 R.J. Glauber, J. Math. Phys. 4, 294 (1963).


\bibitem{c_ring}
C. Monthus, J. Stat. Mech. (2019) 023206

\bibitem{c_interactions}
C. Monthus, J. Phys. A: Math. Theor. 52, 135003 (2019)


\bibitem{c_open}
C. Monthus, J. Phys. A: Math. Theor. 52, 025001 (2019)


\bibitem{singh} 
N. Singh and and B. Wynants, J. Stat. Mech. P03007 (2010)

\bibitem{bertini}
L. Bertini, R. Ch\'etrite, A. Faggionato and D. Gabrielli, 
Annales Henri Poincar\'e  19, 3197 (2018).


\bibitem{barato17}
A. C. Barato and R. Ch\'etrite, J. Stat. Mech. 053207 (2018).


\bibitem{barato18}
A. C. Barato, R. Ch\'etrite, A. Faggionato and D. Gabrielli,
New J. Phys. 20, 103023 (2018).







\end{thebibliography}
\end{document}